\documentclass[ reprint,
nofootinbib,
 amsmath,amssymb,prd,
 aps,onecolumn]{revtex4-2}

\usepackage{color}
\usepackage{dutchcal}
\usepackage{graphicx}
\usepackage{dcolumn}
\usepackage{bm}

\usepackage{multirow}

\begin{document}

\title{Dynamical perturbations around an extreme mass ratio inspiral near resonance}

\author{Makana Silva}
 \email{silva.179@osu.edu}
\affiliation{Center for Cosmology and AstroParticle Physics (CCAPP)}

\author{Christopher Hirata}
\email{hirata.10@osu.edu}
\affiliation{Center for Cosmology and AstroParticle Physics (CCAPP)}

\date{July 15, 2022}

\begin{abstract}
Extreme mass ratio inspirals (EMRIs) -- systems with a compact object orbiting a much more massive (e.g., galactic center) black hole -- are of interest both as a new probe of the environments of galactic nuclei, and their waveforms are a precision test of the Kerr metric.
This work focuses on the effects of an external perturbation due to a third body around an EMRI system. This perturbation will affect the orbit most significantly when the inner body crosses a resonance with the outer body, and result in a change of the conserved quantities (energy, angular momentum, and Carter constant) or equivalently of the actions, which results in a subsequent phase shift of the waveform that builds up over time.
We present a general method for calculating the changes in action during a resonance crossing, valid for generic orbits in the Kerr spacetime. We show that these changes are related to the gravitational waveforms emitted by the two bodies (quantified by the amplitudes of the Weyl scalar $\psi_4$ at the horizon and at $\infty$) at the frequency corresponding to the resonance.
This allows us to compute changes in the action variables for each body, without directly computing the explicit metric perturbations, and therefore we can carry out the computation by calling an existing black hole perturbation theory code.
We show that our calculation can probe resonant interactions in both the static and dynamical limit. We plan to use this technique for future investigations of third-body effects in EMRIs and their potential impact on waveforms for LISA. 
\end{abstract}

\maketitle

\section{Introduction} \label{sec:intro}

Gravitational wave (GW) astronomy has been a rapidly growing subject, with the development of new GW detectors providing new probes for understanding the fundamental properties of gravity from both an astrophysical and cosmological perspective \citep{2013arXiv1305.5720E}. Gravitational waves are the propagation of spacetime distortions generated by massive objects as they accelerate through the Universe. GW astronomy provides a new probe for observing astrophysical phenomena in our Universe which, in parallel with electromagnetic observations, opens a new avenue of answering some of the most compelling mysteries about our Universe such as the Hubble tension \cite{2021ApJ...909..218A}, black hole spacetime dynamics \citep{2010ApJ...718..739M, 2019A&A...627A..92G}, and inflation models \citep{2022arXiv220313842B}, just to name a few. The current and future ground-based GW detectors, such as LIGO \citep{2019BAAS...51g..35R}, VIRGO \citep{2012JInst...7.3012A}, KAGRA \citep{2021PTEP.2021eA103A}, and Cosmic Explorer \citep{2019BAAS...51g..35R} can probe the $\sim100\,$Hz frequency range, corresponding to emission of GWs from compact binary inspirals \cite{2017PhRvL.119p1101A}. The next generation of detectors, such as LISA \citep{2013arXiv1305.5720E}, will open the possibility to observe in the lower frequency range ($\sim 10^{-3}$~Hz), sourced by inspirals near super massive black holes and the stochastic GW background \citep{2019arXiv190403187T}. Thus, in order to extrapolate physical information in this lower range, we will need analytic/computational methods for constructing the appropriate waveforms in this regime.

One source of low frequency GWs are from Extreme Mass Ratio Inspirals (EMRIs), which are the result of two compact objects with a mass ratio very far from unity ($\lesssim 10^{-4}$) orbiting around each other and spiraling in due to radiation reaction \citep{2011PhRvD..83h4023B, 2019RPPh...82a6904B}. In a typical astrophysical context, the smaller body could be a neutron star or stellar mass black hole, and the larger body would be a massive or supermassive black hole. The frequencies from EMRIs are determined by the more massive body and fall around the mHz range for massive black holes, making them ideal targets for space-based surveys such as LISA \citep{2013arXiv1305.5720E}. EMRIs are of great interest for several reasons. On the {\em fundamental physics} side, we can use EMRIs as a test of GR: they are as close as we can come to probing the dynamics of test particles around black hole spacetimes (Schwarszchild or Kerr); and because they survive for many orbital cycles in the strong-field regime, their waveforms are sensitive to even small deviations from the Kerr metric \citep{2007PhRvD..75d2003B, 2019A&A...627A..92G}. On the {\em astrophysical} side, we expect that the dense stellar environment of galactic center black holes should also contain stellar remnants that lead to EMRIs; the rate and parameters of EMRIs provide an observational window into this environment \citep{2010ApJ...718..739M, 2018MNRAS.477.4423A}. 
Realistically, we would expect that most galactic center environments would not consist of a single compact object, but rather, a collection of compact objects \citep{2010ApJ...718..739M}; some models predict distances of order $100M$ to the next-nearest stellar-mass black hole that will inspiral \citep{2019PhRvL.123j1103B}. This motivates the need to study not only the dynamics of EMRIs, but also their interactions under the influence of nearby external bodies. From the theorist's point of view, this means we should investigate the hierarchical three-body problem in general relativity.

A third body introduces the dynamical effect of resonances and allows for the case of studying the changes to gravitational wave emissions during a resonance crossing. Several studies have already shown the impact resonances would have on the inspiral time and phase of the orbit that are not accounted for in standard post-Newtonian approximations \citep{2012PhRvL.109g1102F,2011MNRAS.414.3198H,2011PhRvD..83j4024H}. When the two bodies cross a resonance, changes in their ``constants'' of motion due to the perturbation from the other bodies do not average to zero over an orbit; hence, over many orbits, there can be a significant change in the orbital elements of the system. In the general case, the three-body problem (both in the non and relativistic regimes) leads to chaotic behaviours of motion \citep{Wardell:2002iq} and has led to the development of different approximation schemes for describing multi-body dynamics in GR e.g parameterized post-Minkowskian/Newtonian expansion \citep{2004PhR...400..209K, 2010ApJ...724..39M, 2021PhRvD.103f4010L}, and the static-limit for perturbations around a two body system using explicit metric perturbation calculations \citep{2019PhRvL.123j1103B, 2022arXiv220504808G}. But in the extreme mass ratio regime, we can analytically track the evolution of the inspiral using black hole perturbation theory, which differs from the usual techniques used to study resonant interactions as it provides an analytic method for probing the strong field regime of the inspiral under the influence of a perturbation. 

If we are in an extreme mass ratio regime but the inner body is in a highly relativistic orbit ($r/M$ not very large), the most useful analytic tool is black hole perturbation theory (BHPT). BHPT is a perturbative expansion of the Einstein field equations in orders of the mass ratio. The perturbation equations around a spinning black hole background spacetime can be separated into radial and angular ordinary differential equations \cite{1973ApJ...185..635T}. BHPT has been applied to problems involving both freely propagating gravitational degrees of freedom (e.g., scattering of gravitational waves, stability of the Kerr black hole, gravitational wave Hawking radiation) \cite{1973ApJ...185..649P, 1974ApJ...193..443T, 1976PhRvD..14.3260P, 1978ApJS...36..451M, 2016JCAP...10..034D, 2019EPJC...79..693A, 2021hgwa.bookE..42L}, and to problems with sources, e.g., to compute the waveforms produced by particles orbiting near a spinning black hole \citep{1993PThPh..90..595S, 2001PhRvD..64f4004H, 2002PhRvD..66d4002G, 2006PhRvD..73b4027D}.

Our analysis in this paper deals with an external perturbing body in orbit around an EMRI. We will quantify the effects of this perturber on the orbital dynamics of an inner body inspiralling into a spinning black hole (Kerr spacetime) by looking at the change in the action variables of the inner orbit. These changes are directly related to the change in fundamental frequencies and hence the phase of the emitted waveforms \citep{2000PhRvD..61h4004H, 2001PhRvD..64f4004H}. The phase is the quantity that can be measured most precisely from a signal that is many cycles long, and has thus attracted attention for tests for general relativity \citep{2013arXiv1305.5720E}. In this study, we will relate the change in the action variables to the energy transfer between the EMRI and perturber during a resonance crossing. This transfer of energy can be calculated by looking at the superposition of the gravitational perturbations generated by the EMRI and perturber, which is contained in the gauge-invariant Weyl curvature scalar, $\psi_{4}$ \citep{1967JMP.....8..265B,1973ApJ...185..635T}. This Weyl scalar satisfies a separable partial differential equation on the Kerr background spacetime, and allows us to compute some types of multi-body interactions without computing metric perturbations \citep{2011MNRAS.414.3212H}. It is also related to the emitted waveform, so there are many existing codes that can compute the amplitude of $\psi_4$ sourced by a small body orbiting in the Kerr spacetime \citep{2011MNRAS.414.3212H, 2011PhRvD..83j4024H, 2019PhRvL.123j1103B, 2022arXiv220504808G}.

Our paper is organized as follows. Section~\ref{sec:formalism} reviews the relevant formalism and tools used in this analysis. Section~\ref{sec: change_in_action} begins the analysis with computing the change in the action variables and establishing the relevant quantities we need to compute, followed by laying out the calculations to the energy transfer near a resonance crossing, which is then related to the amplitudes of the bodies' gravitational waveforms. Section~\ref{sec: Testing_models} compares our calculation to other results in the literature in the cases where they overlap, such as comparing the torque induced during a precession resonance crossing in the Keplerian regime and the tidal torque due to a tidal resonance crossing from a static perturber. Section~\ref{section: discussion} discusses the results and implications of this analysis.  Appendix~\ref{app:time} details the difference between orbit-averaged quantities of evolution with respect to proper time and coordinate time (which we use for evolving quantities).

\section{Formalism and conventions} \label{sec:formalism}

This section gives an overview of the ``toolbox'' used for our computation. We will be working in units with $G=c=1$.

\subsection{The particles and orbits in the Kerr spacetime}

Here we are studying a three body system: a central black hole (provides the Kerr geometry), and two orbiting bodies. We denote the central black hole's mass by $M$ and its specific angular momentum by $a$ (with dimensionless spin parameter $a_\star=a/M)$. The masses of the inner and outer bodies are denoted by $\mu_{\rm inner}$ and $\mu_{\rm outer}$, respectively.

We work in the Boyer-Lindquist \cite{1967JMP.....8..265B} coordinate system $(t,r,\theta,\phi)$, with the $-+++$ signature.
Four-vector quantities will be denoted with a boldface print ${\bf J}$ with components indexed by Greek letters. Quantities with a tilde $\widetilde{A} \equiv A/\mu_{A}$ have the mass of the relevant particle divided out, so for example one may write the 4-velocity of one of the orbiting bodies as as $\widetilde p_\alpha = p_\alpha/\mu = u_\alpha$, where ${\bf p}$ is the 4-momentum.

We parameterize particle trajectories in this paper by $x^i(t)$, where $i=1,2,3$ are the spatial indices and $t$ is the Boyer-Lindquist coordinate time. This is distinct from other possible choices of parameter, such as the proper time $\tau$ (related by $dt/d\tau = u^t$) or the Mino \cite{2003PhRvD..67h4027M} time variable $\lambda$ (related by $dt/d\lambda = u^t\Sigma$, where $\Sigma = r^2 + a^2\cos^2\theta$). The use of the coordinate time as a parameter, following, e.g.,  Refs.~\cite{2011MNRAS.414.3198H,2011MNRAS.414.3212H}, is a convenient choice for the canonical action-angle variable formalism, and simplifies the interaction between the particles since all calculations can be in terms of the same time variable. It also simplifies the description for what a distant observer (e.g., LISA) will see, since $t$ is more easily related to the observer's clock than $\tau$ or $\lambda$. Furthermore, the Hamiltonian constructed using Boyer-Lindquist coordinate time corresponds to the conserved energy that results from the timelike Killing vector with $\xi^t=1$ and $\xi^i=0$.
Dots in this paper always represent derivatives with respect to $t$.

If external perturbations and the self-force are neglected, then the orbiting bodies will travel on geodesics. Geodesics in the Kerr spacetime can be described with the three positions $x^i(t)$ and three conjugate momenta $p_i(t)$. They have three involute conserved quantities: the energy per unit mass $\tilde{\mathcal E}= -u_t$ (which is related to $u_i$ by the mass shell relation $g^{\mu\nu}u_\mu u_\nu = -1$); the angular momentum per unit mass $\tilde{\mathcal L} = u_\phi$; and the Carter constant $\tilde{\mathcal Q}$ \cite{1968PhRv..174.1559C}. As such, one can construct action-angle variables: we have three actions $(J_r,J_\theta,J_\phi)$, and three conjugate angles $(\psi^r,\psi^\theta,\psi^\phi)$. It is convenient to write the actions in terms of action per unit mass $\tilde J_i$. For an unperturbed orbit, we have
\begin{equation}
\tilde J_i = {\rm constant},
~~~
\tilde\psi^i = {\rm constant} + \Omega^it,
~~
\Omega^i = \frac{\partial H}{\partial J_i} = \frac{\partial \tilde H}{\partial \tilde J_i},
\label{eq:geodesic}
\end{equation}
where $H=-p_t$ is the single-particle Hamiltonian, and
$\Omega^i$ are the fundamental angular frequencies for each of the three directions in the torus ($i\in\{r,\theta,\phi\}$).
There is a choice of origin for the angles. Following Ref.~\cite{2011MNRAS.414.3212H}, we place $\psi^i=0$ at pericenter ($r=r_{\rm min}$), at the ascending node ($\theta=\pi/2$, $u^\theta<0)$, and at the prime meridian ($\phi=0$).
The actions and angles constructed here are thus the same as in Ref.~\cite{2011MNRAS.414.3212H}. In what follows, 3-vectors in action-angle space will be represented by a $\vec{J}$ and indexed by Latin letters.

The actions are equivalent to those of Flanagan \& Hinderer \cite{2012PhRvL.109g1102F}, but the angles are different because the choice of time parameter is different.

\subsection{Self-force and resonance contributions to the orbital evolution}

We now focus on the evolution of the inner body. There are two important corrections to Eq.~(\ref{eq:geodesic}) in this problem. Since the inner body has a non-zero rest mass, it will experience an interaction with its own gravitational perturbation (\textit{self-force}). The dissipative part of the self-force causes a change in the actions and corresponds to emission of gravitational waves. 

The outer body also acts on the inner body: it produces an external \textit{tidal field} that could perturb both the actions and angles of the inner body.
This gives us the usual equations of motion for the test particle with contributions due to the self-force interactions and the external perturbing field:
\begin{equation}
    \dot{\vec{\psi}}_{\rm inner} \approx \vec{\Omega}_{\rm inner}(\tilde{\vec J}_{\rm inner}) + \mu_{\rm outer} \vec{g}_{\rm td}(\vec{\psi}_{\rm inner},\tilde{\vec J}_{\rm inner},\vec{\psi}_{\rm outer},\tilde{\vec J}_{\rm outer}) + \mu_{\rm inner} \vec{g}_{\rm sf}(\vec{\psi}_{\rm inner},\tilde{\vec J}_{\rm inner})
\label{eq: psi_dot}
\end{equation}
and
\begin{equation}
   \dot{\tilde{\vec{J}}}_{\rm inner} \approx  \tilde{\vec{G}}_{\rm td}(\vec{\psi}_{\rm inner},\tilde{\vec J}_{\rm inner},\vec{\psi}_{\rm outer},\tilde{\vec J}_{\rm outer}) + \tilde{\vec{G}}_{\rm sf}(\vec{\psi}_{\rm inner},\tilde{\vec J}_{\rm inner}),
\label{eq: EOM}
\end{equation}
where $g$ and $G$ represent the perturbative changes in the angle and action variables due to external perturber (subscript ``td'') and the self-force reaction on the orbiting mass due to emission of radiation (subscript ``sf''). 

\begin{figure}
    \centering
    \includegraphics[scale=0.5]{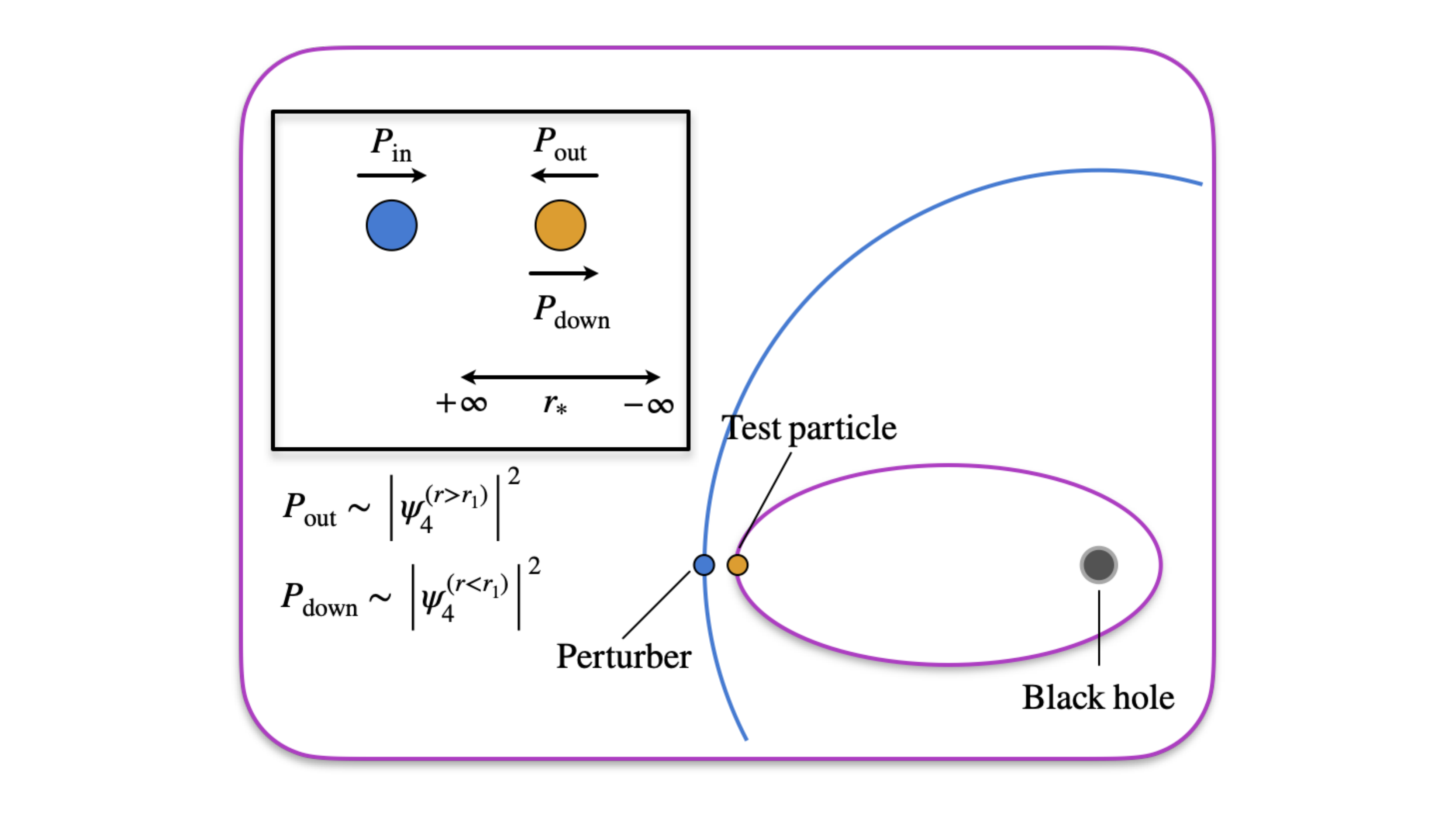}
    \caption{A model of our three body system. The test particle (purple) can take any generic geodesic while the external perturber (blue) is assumed to be on an equatorial, circular orbit. The test particle will follow some inspiral trajectory due to the emission of gravitational radiation. The presence of the external perturber will affect the emitted radiation from the test particle by its own gravitational field. The power of the emitted radiation can be computed in the regions between the particle and the BH ($r < r_1$) and the  particle  infinity ($r > r_1$), where $r_1$ is the location of the perturbing body. The $\psi_ 4$ scalar for both regions will have contributions from the pure gravitational radiation due to the perturber.} 
    \label{fig:Kerr_model}
\end{figure}

The motion of each body has a set of frequencies that are integer linear combinations of the fundamental frequencies, $\omega = \vec N\cdot\vec\Omega$, where $\vec N$ has integer components: $\vec{N} = (n,k,m)$ or $\omega = n\Omega^r + k\Omega^\theta + m\Omega^\phi$. A resonance may occur when the inner and outer body share a frequency in common. We define the resonance condition
\begin{equation}
\label{eq:resonance_condition}
\Delta \omega \equiv \vec{N}_{\rm outer} \cdot \vec{\Omega}_{\rm outer} - \vec{N}_{\rm inner} \cdot \vec{\Omega}_{\rm inner} = 0.
\end{equation}
and the resonant angle
\begin{equation}
\label{eq:resonance_angle}
\Theta \equiv \vec{N}_{\rm outer} \cdot \vec{\psi}_{\rm outer} - \vec{N}_{\rm inner} \cdot \vec{\psi}_{\rm inner},
\end{equation}
which is stationary ($\dot\Theta=0$) at resonance.

Because of the symmetries of the Kerr spacetime, not every resonance has a physical effect: azimuthal symmetry and reflection symmetry across the Equator suggest the ``selection rules'' $m_{\rm inner}=m_{\rm outer}$ and $k_{\rm inner}-k_{\rm outer} =$\,even, respectively. In Section~\ref{ss:selection}, we will see this by explicit calculation that this is the case.

While the resonance formalism is general, the main application in this paper is to generic orbits for the inner body and circular, equatorial orbits for the outer perturber. For circular orbits, the particle position and momentum do not depend on $\psi^r$ and so there is a selection rule $n_{\rm outer}=0$; and similarly for equatorial orbits there is a selection rule $k_{\rm outer}=0$. In this case, we can expand our resonance condition to be
\begin{equation}
\label{res_condition}
n_{\rm inner}\Omega^{r}_{\rm inner} + k_{\rm inner}\Omega^{\theta}_{\rm inner} + m\Omega^{\phi}_{\rm inner} = m\Omega^{\phi}_{\rm outer},
\end{equation}
where for prograde orbits
\begin{equation}
\label{eq:angularfreq_Kerr}
    \Omega^{\phi}_{\rm outer} = \frac{M^{1/2}r^{-3/2}}{1 + aM^{1/2}r^{-3/2}}.
\end{equation}

\subsection{Gravitational perturbations}

The quantity of interest that encodes the information of the gravitational radiation is the Weyl scalar, $\psi_4(r,\theta,\phi,t)$ \cite{1962JMP.....3..566N,1973ApJ...185..635T}. This function is a radial-polar decomposition of the solutions to the ``master equation'' (Eq.~4.7 in \citep{1973ApJ...185..635T}) for field quantities with spin weight $s=\pm2$. This is a result of the separation of this master equation into radial and angular parts and the form of $\psi_{4}$ as
\begin{eqnarray}
\label{eq: psi_4}
\psi_4(r, \theta, \phi, t) = (r - ia\cos{\theta})^{-4}\int^{\infty}_{-\infty} \frac{d\omega}{2\pi} \sum_{lm} \mathcal{R}_{lm\omega}(r){}_{-2}S^{a\omega}_{lm}(\theta)e^{i(m\phi - \omega t)},
\end{eqnarray}
where $\omega$ is the frequency of the radiation, $\mathcal{R}_{lm\omega}(r)$ are the radial functions that satisfies the Teukolsky equations, ${}_{-2}S^{a\omega}_{lm}(\theta)$ are the latitude functions that satisfy the angular part of the master equation, and $a$ is the spin parameter ($0<a<M$). It is also useful to convert the typical radial coordinate to the ``tortoise coordinate''
\begin{eqnarray}
\label{eq: r_tort}
r_* = r + \frac{r_{\rm H+}}{\sqrt{1-a^2}}\ln{\frac{r - r_{\rm H+}}{2M}} + \frac{r_{\rm H-}}{\sqrt{1-a^2}}\ln{\frac{r - r_{\rm H-}}{2M}},
\end{eqnarray}
where $r_{\rm H\pm}$ are the horizons, i.e., solutions to $r^2 - 2aM +a^2=0$. This choice of radial coordinate makes the numerical implementation of what is to come next much simpler since the horizon, $r_{\rm H+}$, gets mapped to $-\infty$ while getting much farther from the black hole retains its usual definition.

The radial functions can be computed using a Green's function method to the inhomogeneous ordinary differential equation\footnote{We suppressed the indices ${lm\omega}$ for brevity.} for a particle emitting gravitational radiation towards the black hole and infinity
\begin{eqnarray}
\label{eq: Rad_EQU}
\Delta^2\frac{d}{dr}\left( \Delta^{-1} \frac{d\mathcal{R}}{dr} \right) - V\mathcal{R} = \mathcal{T},
\nonumber\\
V = -\frac{K^2 + 4i(r - M)K}{\Delta} + 8i\omega r + \widetilde{\mathcal{E}} - 2am\omega + a^2\omega^2 - 2,
\end{eqnarray}
where $K = \omega(r^2 + a^2) - am$, $\Delta = r^2 - 2aM +a^2$, and $\mathcal{T}$ is the source function defined in Eqs.~(71--73) of Ref.~\cite{2011MNRAS.414.3212H}.
The radial solutions interior to the pericenter $r_{\rm min}$ and exterior to the apocenter $r_{\rm max}$ are:
\begin{eqnarray}
\label{eq: rad_soln}
\mathcal{R}_{lm\omega} = \sum_{\vec{N}}Z^{\rm down}_{lm,\vec{N}}\mathcal{R}_{1}(r)e^{-i\vec{N} \cdot \vec{\psi}^{0}}2\pi \delta(\omega - \vec{N} \cdot \Omega) ~~~~{\rm at}~r<r_{\rm min},
\nonumber\\
Z^{\rm down}_{lm,\vec{N}} = \frac{\Delta}{W_{13}}\int \frac{d^3\vec\psi}{(2\pi)^3} \left( \mathcal{A}_{0}\mathcal{R}_{3} - \mathcal{A}_{1}\mathcal{R}'_{3} + \mathcal{A}_{2}\mathcal{R}''_{3} \right)e^{i\vec{N} \cdot \vec{\psi}},
\end{eqnarray}
and
\begin{eqnarray}
\label{eq: rad_soln2}
\mathcal{R}_{lm\omega} = \sum_{\vec{N}}Z^{\rm out}_{lm,\vec{N}}\mathcal{R}_{3}(r)e^{-i\vec{N} \cdot \vec{\psi}^{0}}2\pi \delta(\omega - \vec{N} \cdot \Omega) ~~~~{\rm at}~r>r_{\rm max},
\nonumber\\
Z^{\rm out}_{lm,\vec{N}} = \frac{\Delta}{W_{13}}\int \frac{d^3\vec\psi}{(2\pi)^3} \left( \mathcal{A}_{0}\mathcal{R}_{1} - \mathcal{A}_{1}\mathcal{R}'_{1} + \mathcal{A}_{2}\mathcal{R}''_{1} \right)e^{i\vec{N} \cdot \vec{\psi}},
\end{eqnarray}
where $Z^{\rm down,out}$ are the amplitudes for the ingoing/outgoing waves, $\mathcal{R}_{1,3}(r)$ are the asymptotic solutions for an ingoing/outgoing wave, respectively, $W_{13}$ is the Wronskian for a given $R_{1,3}$, and $\mathcal{A}_{0,1,2}$ are functions of the weighted spheroidal harmonics given in \citep{2000PhRvD..61h4004H,2011MNRAS.414.3212H}.

The $Z^{\rm down,out}$ obey a selection rule that $m=N_\phi$ (see Section 4.3 of Ref.~\cite{2011MNRAS.414.3212H}). There is also a symmetry relation resulting from reflection across the equator: 
if we increment $\psi^\theta \rightarrow \psi^\theta + \pi$, then the particle's trajectory is flipped across the equator ($\theta\rightarrow\pi-\theta$, $u_\theta\rightarrow-u_\theta$, other coordinates unchanged).\footnote{Flipping the particle's trajectory across the equator is a canonical transformation. We also know that the $\tilde{J}_{\rm i}$ are functions of $(\tilde{\mathcal{E}}, \tilde{\mathcal{Q}}, \tilde{\mathcal{L}})$, which are adiabatic invariants of the Kerr system, so they do not change under this transformation. Therefore, we know by the direct conditions (see, e.g., Eq.~9.48a of Ref.~\cite{2002clme.book.....G}) that the transformed angles are related to the original angles by $\partial\psi'^{i}/\partial\psi^j = \partial\tilde J_j/\partial\tilde J_i = \delta^i_j$, and so the change in the angles go as $\psi'{^i} = \psi^i + f^i(\tilde{J}_{\rm i})$. And since applying this flip twice should return the particle to its original trajectory, then we know either $f^i = 0$ or $\pi$ for each component $i\in\{r,\theta,\phi\}$. Inspection of the Keplerian limit allows us to discern that the change is 0 for $\psi^r$ and $\psi^\phi$ and $\pi$ for $\psi^\theta$. Thus, flipping the particle's trajectory across the equator is equivalent to incrementing in the angle coordinate $\psi^\theta$ by $\pi$.}
The function $_{-2}S^{a\omega}_{lm}(\theta,\phi)$ obeys the same reflection rule as the spin-weighted spherical harmonics,
\begin{equation}
_{-2}S^{a(-\omega)}_{l,-m}(\pi-\theta,\phi)
=
(-1)^l
{_{-2}}S^{a\omega\,\ast}_{lm}(\theta,\phi).
\end{equation}
By following each step in Section 4.3 of Ref.~\cite{2011MNRAS.414.3212H}, we can see that under the transformation $\vec N\rightarrow -\vec N$ and $\psi^\theta\rightarrow \psi^\theta+\pi$, one sees that the radial functions $\mathcal R_{1,3}$ pick up a complex conjugate, the sources $\mathcal A_{0,1,2}$ pick up a complex conjugate and a factor of $(-1)^l$, and the factor $e^{i\vec N\cdot\vec\psi}$ picks up a complex conjugate and a factor of $e^{iN_\theta\pi} = (-1)^{N_\theta}$. So overall, we conclude that
\begin{equation}
Z^{\rm down,out}_{l,-m,-\vec N}
= (-1)^{l+N_\theta} Z^{{\rm down,out}\,\ast}_{l,m,\vec N}.
\label{eq:equator}
\end{equation}

The $Z^{\rm down/out}$ amplitudes and appropriate scattering coefficients were computed using the code developed in Ref.~\cite{2011MNRAS.414.3212H}, which is a {\tt C} implementation of a Teukolsky equation solver. 

\section{Calculation of the change in the actions}\label{sec: change_in_action}

We can now compute the rate of change in the energy and the action variables of the inner body due to perturbation from the outer body. Our method is similar to that of Ref.~\cite{2011MNRAS.414.3212H}: we will replace the outer body with a gravitational wave coming in from $\infty$ that produces the same $\psi_4$ in the region where the inner body is located. The outgoing radiation (to $\infty$) and downgoing radiation (to $r_{\rm H+}$) are superpositions of this incident radiation (with appropriate scattering coefficients) and the radiation emitted by the inner body. We can then use conservation of energy to determine how much power is absorbed by the inner body, and then knowledge of the integers in the resonance enables us to determine $\dot{\vec J}_{\rm inner}$.

The Weyl scalar perturbation generated by the outer body in the region of space $r<r_{\rm min,outer}$ is given by Eq.~(\ref{eq: rad_soln}). This is of course a vacuum solution to the Teukolsky equation. We focus on the frequency modes with $\omega\neq 0$ (the $\omega=0$ terms require special treatment, but in any case result in no energy transfer because they are time-independent). We then replace the outer perturber with an incoming gravitational wave from $\infty$ with the same $\psi_4$ at $r<r_{\rm min,outer}$. This replacement does not change the spacetime at $r<r_{\rm min,outer}$, because there exists a procedure to reconstruct the metric (modulo gauge transformations) once $\psi_4$ is determined \cite{1975PhRvD..11.2042C, 1978PhRvL..41..203W, 2003PhRvD..67l4010O}. It is not necessary in this paper to actually carry out this reconstruction procedure.

Our method can be seen graphically in Fig.~\ref{fig:Kerr_model} by removing the blue (outer) body and replacing it with a gravitational wave. This will effectively turn this into a 2-body problem (black hole + inner body) but with a different boundary condition from the usual case (since now we have an incoming gravitational wave). We can compute the absorbed power by the inner body using conservation of energy.

\subsection{Relation of the change in action to resonant energy transfer}

In order to quantify the changes to the action variables of the orbiting particle, we will apply the stationary phase method on the unperturbed trajectory (i.e., the Born approximation treating $\mu_{\rm outer}$ as a perturbation). This approach was presented in the discussion of Ref.~\cite{2012PhRvL.109g1102F} (for the transient resonances where $\Omega^\theta:\Omega^r$ is rational) and Eq.~(7) of Ref.~\cite{2019PhRvL.123j1103B} (for tidal resonances). The differences in our case do not substantially change the derivation: they are (i) that we work with Boyer-Lindquist time $t$ instead of Mino time as the independent variable; and (ii) the frequency derivatives include the outer body as well.

We want the change in $J_{{\rm inner},i}$ due to the external perturber's gravitational presence. Integrating over some range of time $t_1$ to $t_2$, we may write:
\begin{eqnarray}
\label{eq: delta_J}
\Delta \tilde{J}_{{\rm inner},i} &=& \int_{t_1}^{t_2} \tilde{G}_{\rm td,i}(\vec{\psi}_{\rm inner},\tilde{\vec{J}}_{\rm inner},
\vec{\psi}_{\rm outer},\tilde{\vec{J}}_{\rm outer})\, dt
\nonumber \\
&=& \int_{t_1}^{t_2} \sum_{\vec{N}_{\rm inner},\vec N_{\rm outer}} \tilde{G}_{\vec{N}_{\rm inner},\vec N_{\rm outer}, {\rm td},i}(\tilde{\vec{J}}_{\rm inner}, \tilde{\vec J}_{\rm outer})e^{i(\vec{N}_{\rm outer} \cdot \vec{\psi}_{\rm outer} - \vec{N}_{\rm inner} \cdot \vec{\psi}_{\rm inner})}\, dt,
\end{eqnarray}
where we expanded the tidal contribution to the action variables in a Fourier series over the angles for both the inner and outer body orbits. If the frequencies evolve slowly, then the largest contribution to this integral will come when the particle and perturber are in resonance, that is, when the angle $\Theta = \vec{N}_{\rm outer} \cdot \vec{\psi}_{\rm outer} - \vec{N}_{\rm inner} \cdot \vec{\psi}_{\rm inner}$ defined by Eq.~(\ref{eq:resonance_angle}) is stationary ($\dot\Theta=0$ at $t=t_{\rm res}$). We can then expand
\begin{equation}
\Theta = \Theta_{\rm res} + \frac12\Gamma (t-t_{\rm res})^2 + ...\,,
\end{equation}
where $\Theta_{\rm res}$ is the value of the resonant angle at resonance crossing $t=t_{\rm res}$; and $\Gamma$ is the rate of change of the frequency,
\begin{equation}
\Gamma \equiv \Delta \dot{\omega} = \vec{N}_{\rm outer} \cdot \dot{\vec{\Omega}}_{\rm outer}  - \vec{N}_{\rm inner} \cdot \dot{\vec{\Omega}}_{\rm inner} 
 = {N}_{{\rm outer},j} \frac{\partial {\Omega}^j_{\rm outer}}{\partial \tilde{J}_{{\rm outer},i}}\dot{\tilde{J}}_{{\rm sf,outer},i} - {N}_{{\rm inner},j} \frac{\partial \Omega^j_{\rm inner}}{\partial \tilde{J}_{{\rm inner},i}}\dot{\tilde{J}}_{{\rm sf,inner},i}.
\label{eq:Gamma0}
\end{equation}
The integral in Eq.~(\ref{eq: delta_J}) then picks up most of its contribution from a time interval $|t-t_{\rm res}|\lesssim {\rm few}\times|\Gamma|^{-1/2}$; before and after this, the integrand is rapidly oscillating. The total change in action over a time interval from well before resonance crossing to well after is then obtained by taking only the terms that are resonant ($\sum_{\rm res}$); taking the limits of integration to be $-\infty$ to $\infty$; and pulling $\tilde{G}_{\vec{N}_{\rm inner},\vec N_{\rm outer}, {\rm td},i}(\tilde{\vec{J}}_{\rm inner}, \tilde{\vec J}_{\rm outer})$ outside the integral:
\begin{align}
\label{eq: delta_J_expand}
    \Delta \tilde{J}_{\rm inner,i} \approx \sum_{\rm res}  \tilde{G}_{\vec{N}_{\rm inner},\vec N_{\rm outer}, {\rm td},i}(\tilde{\vec{J}}_{\rm inner}, \tilde{\vec J}_{\rm outer})
    \int_{-\infty}^{\infty} e^{i(\Theta_{\rm res} + \frac12\Gamma(t-t_{\rm res})^2)} dt
    \nonumber\\
    = \sum_{\rm res}  \tilde{G}_{\vec{N}_{\rm inner},\vec N_{\rm outer}, {\rm td},i}(\tilde{\vec{J}}_{\rm inner}, \tilde{\vec J}_{\rm outer}) e^{i\Theta_{\rm res}}
    e^{i\pi ({\rm sgn}\,\Gamma)/4} \sqrt{\frac{2 \pi}{|\Gamma|}}.
\end{align}
We have left the summation over resonances here, since where there is one resonant term there will be many (at the very least, if there is an $\vec N_{\rm inner}:\vec N_{\rm outer}$ resonance, any integer multiple $s\vec N_{\rm inner}:s\vec N_{\rm outer}$ with $s\in{\mathbb Z}$ will also be resonant).

We see that Eq.~(\ref{eq: delta_J_expand}) depends on both the change in action variables due to the external perturber's tidal field (via $\tilde{G}_{\vec{N}_{\rm inner},\vec N_{\rm outer}, {\rm td},i}$) and both bodies' self-force reaction ($\dot {\tilde J}_{\rm sf,i}$ in Eq.~\ref{eq:Gamma0} for $\Gamma$).
The expression for change in action due to the self force, $J_{\rm sf,i}$ is given in Eq.~(12) of Ref.~\cite{2019PhRvL.123j1103B} or following the logic in Eqs.~(22--24) of Ref.~\cite{2011PhRvD..83j4024H}:
\begin{equation}
\label{eq: J_dot_sf}
    \widetilde{\dot{J}}_{{\rm sf},i} = - \sum_{\vec{N}} \sum_{ lm} \frac{\mu N_{ i}}{2\omega^3} \left( |\widetilde{Z}^{\rm out}_{\vec{N}, lm}|^2 + \alpha_{ lm} |\widetilde{Z}^{\rm down}_{\vec{N}, lm}|^2 \right),
\end{equation}
where the sum is over all the modes and the angular quantum number $l$; $\alpha_{ lm}$ is the normalization factor for energy absorbed by the black hole \citep{1974ApJ...193..443T, 2000PhRvD..61h4004H, 2006PhRvD..73b4027D}; and the equation can be applied to either the inner or outer body. This sum is made over all modes, since they all contribute to the self-force, regardless of resonance conditions.

We now need to compute the rate of change of the action variables of the inner body due to the external body, i.e., we need $\tilde{G}_{\vec{N}_{\rm inner},\vec N_{\rm outer}, {\rm td},i}$. Our next step is to relate this to the resonant energy transfer.

\subsection{The changes in the actions in terms of the resonant energy transfer}

The key to relating the actions to the energy transfer is that -- for a given resonance -- only certain linear combinations of the actions can change \cite{1978mmcm.book.....A}. The motion of the inner body in the spacetime of the black hole and with the outer body as a perturbation is described by a Hamiltonian, with a perturbation $\Delta{\mathcal H}$. (We won't actually need to compute the form of $\Delta{\mathcal H}$.) The rates of change of the action variable of the inner body due to the external perturber are then given by Hamilton's equations:
\begin{equation}
\langle \dot {\tilde J}_{\rm td,i} \rangle = -\left\langle \frac{\partial \Delta\tilde{\mathcal{H}}}{\partial \psi^{\rm i}_{\rm inner}}\right\rangle,
\end{equation}
where the brackets denote an orbit average (and not a full torus average in the case where there is a resonance).
If we keep only a resonant term with a $\propto e^{-i\vec N_{\rm inner}\cdot\vec\psi_{\rm inner}}$ angular dependence and $\propto e^{-i\omega t}$ time dependence, then the $\psi^i_{\rm inner}$ derivative pulls down a factor of $iN_{{\rm inner},i}$. Therefore, the contribution of each resonance to $\dot {\tilde J}_{\rm td,i}$ is of the form
\begin{equation}
\langle \dot {\tilde J}_{\rm td,i}\rangle = N_{{\rm inner},i}{\mathcal K},
\end{equation}
where the constant ${\mathcal K}$ does not depend on $i$ (i.e., it is the same for $i=r$, $\theta$, or $\phi$).
But the orbit-averaged power absorbed by the inner particle is
\begin{eqnarray}
\label{eq: P_abs}
P_{\rm abs} 
= \left\langle \frac{\partial\Delta{\mathcal H}}{\partial t} \right\rangle
= \mu_{\rm inner} \left\langle \frac{\partial\Delta\tilde {\mathcal H}}{\partial t} \right\rangle
= \mu_{\rm inner} \omega{\mathcal K}.
\end{eqnarray}
(Here the last equality came from the fact that our resonant term has $\propto e^{-i\omega t}$ time dependence.)
This allows us to relate the change in the action due to tidal effects to the power absorbed by the test particle
\begin{equation}
\langle 
\dot{\tilde{J}}_{\rm td,i} \rangle = \frac{ P_{\rm abs}N_{{\rm inner},i}}{ \mu_{\rm inner} \omega}.
\label{eq: P_abs_J_dot}
\end{equation}
We can see that the rate of change of the action variables is proportional to the absorbed power of the inner body. Our next step is to compute this absorbed power.

\subsection{The resonant energy transfer in terms of gravitational waveforms}

We recall that if the radial function near the horizon is a superposition of sinusoidal waves in $t$ so that its Fourier transform is a superposition of $\delta$-functions in $\omega$,
${\cal R}_{l m \omega} = \sum_{\omega_{\rm gw}}
\xi_{lm\omega 1}\mathcal R_1(r) 2\pi\delta(\omega-\omega_{\rm gw})$
(we drop the $\mathcal R_2$ term since no waves are coming ``up'' from the black hole), then the time-averaged power in gravitational waves going down into the black hole is
\begin{equation}
P_{\rm down} = \sum_{\omega_{\rm gw}} \sum_{lm} \frac{\alpha}{2\omega^2}|\xi_{lm\omega 1}|^2
\end{equation}
(this is the usual formula of Ref.~\cite{1974ApJ...193..443T}).
Similarly, near $\infty$, if we have 
${\cal R}_{l m \omega} =\sum_{\omega_{\rm gw}}
[\xi_{lm\omega 3}\mathcal R_3(r) + \xi_{lm\omega 4}\mathcal R_4(r)]2\pi\delta(\omega-\omega_{\rm gw})$, then there is ingoing and outgoing power:
\begin{equation}
P_{\rm in} = \sum_{\omega_{\rm gw}} \sum_{lm} \frac{(2\omega)^8}{2\omega^2|C|^2}|\xi_{lm\omega 4}|^2
~~~{\rm and}~~~
P_{\rm out} = \sum_{\omega_{\rm gw}} \sum_{lm} \frac{1}{2\omega^2}|\xi_{lm\omega 3}|^2.
\end{equation}
In this case, the down-going amplitude is the superposition of that from the inner body and the gravitational wave replacing the outer body:
\begin{equation}
\xi_{lm\omega 1} = \mu_{\rm inner} \sum_{\vec N_{\rm inner}} \tilde Z^{\rm down}_{lm,\vec N_{\rm inner}} e^{-i\vec N_{\rm inner}\cdot\vec\psi^{(0)}_{\rm inner}}
\delta_{\omega,\vec N_{\rm inner}\cdot\vec\Omega_{\rm inner}}
+ \mu_{\rm outer} \sum_{\vec N_{\rm outer}} \tilde Z^{\rm down}_{lm,\vec N_{\rm outer}} e^{-i\vec N_{\rm outer}\cdot\vec\psi^{(0)}_{\rm outer}}
\delta_{\omega,\vec N_{\rm outer}\cdot\vec\Omega_{\rm outer}}.
\label{eq:xi-1}
\end{equation}
(We used a Kronecker delta to ensure that only the terms that emit at frequency $\omega$ are included in the sum; if no resonances are present, then only one term contributes to each frequency.)
For the ingoing and outgoing waves, we use the fact that the radial solutions to the Teukolsky equation near the horizon and $\infty$ are related by a scattering matrix: $\mathcal R_1(r) = c_{13} \mathcal R_3(r) + c_{14}\mathcal R_4(r)$, where the coefficients $c_{13}$ and $c_{14}$ are the scattering matrix elements (which we compute using the code of Ref.~\cite{2011MNRAS.414.3212H}). Then:
\begin{equation}
\xi_{lm\omega 3} = \mu_{\rm inner} \sum_{\vec N_{\rm inner}} \tilde Z^{\rm out}_{lm,\vec N_{\rm inner}} e^{-i\vec N_{\rm inner}\cdot\vec\psi^{(0)}_{\rm inner}}
\delta_{\omega,\vec N_{\rm inner}\cdot\vec\Omega_{\rm inner}}
+ \mu_{\rm outer} \sum_{\vec N_{\rm outer}} c_{13} \tilde Z^{\rm down}_{lm,\vec N_{\rm outer}} e^{-i\vec N_{\rm outer}\cdot\vec\psi^{(0)}_{\rm outer}}
\delta_{\omega,\vec N_{\rm outer}\cdot\vec\Omega_{\rm outer}}
\label{eq:xi-3}
\end{equation}
and
\begin{equation}
\xi_{lm\omega 4} =
 \mu_{\rm outer} \sum_{\vec N_{\rm outer}} c_{14} \tilde Z^{\rm down}_{lm,\vec N_{\rm outer}} e^{-i\vec N_{\rm outer}\cdot\vec\psi^{(0)}_{\rm outer}}
\delta_{\omega,\vec N_{\rm outer}\cdot\vec\Omega_{\rm outer}}.
\label{eq:xi-4}
\end{equation}
Energy conservation then implies that the power transferred to the particle is
\begin{equation}
\dot{\cal E}_{\rm inner} =
P_{\rm in} - P_{\rm out} - P_{\rm down}.
\label{eq:E-inner}
\end{equation}
Note that by replacing the outer body with a gravitational wave, there is no longer a question of how much power is absorbed by the outer body -- it is not there anymore.

In Eq.~(\ref{eq:E-inner}), there are three types of terms: those that contain $\mu_{\rm outer}^2$, $\mu_{\rm inner}^2$, and $\mu_{\rm inner}\mu_{\rm outer}$. The $\mu_{\rm outer}^2$ terms represent the energy balance in gravitational wave scattering off a black hole, and vanish by conservation of energy \cite{1974ApJ...193..443T}. The $\mu_{\rm inner}^2$ terms do not involve the incoming gravitational wave at all -- they represent the power lost by the inner particle due to its self-force. The $\mu_{\rm inner}\mu_{\rm outer}$ terms represent energy transfer to the inner body that depends on the presence of the outer body, and occur only at resonance when there is an $\vec N_{\rm inner}$ and $\vec N_{\rm outer}$ that contribute to the same frequency $\omega$ in Eq.~(\ref{eq:xi-1}) and (\ref{eq:xi-3}).

We now denote by $\sum_{\rm res}$ a summation over resonances where $\omega = \vec N_{\rm inner}\cdot\vec\Omega_{\rm inner} = \vec N_{\rm outer}\cdot\vec\Omega_{\rm outer}$. Then the power absorbed by the inner particle (contribution to $\dot{\cal E}_{\rm inner}$) is
\begin{equation}
\label{eq: power_abs}
P_{\rm abs} = -\left( \delta P_{\rm out} + \delta P_{\rm down} \right),
\end{equation}
where $\delta P_{\rm out}$ comes from the cross terms when we square $\xi_{lm\omega 3}$ (Eq.~\ref{eq:xi-3}):
\begin{equation}
\delta P_{\rm out} = \mu_{\rm inner} \mu_{\rm outer}\sum_{\rm res} \sum_{lm} \Re \left[\frac{c_{13}^\ast\widetilde{Z}^{\rm down\ast}_{lm,\vec{N}_{\rm outer}}\widetilde{Z}^{\rm out}_{lm,\vec{N}_{\rm inner}}}{\omega^2}
e^{i(\vec N_{\rm outer}\cdot\vec\psi^{(0)}_{\rm outer} - \vec N_{\rm inner}\cdot\vec\psi^{(0)}_{\rm inner})}
\right];
\end{equation}
and $\delta P_{\rm down}$ comes from the cross terms when we square $\xi_{lm\omega 1}$ (Eq.~\ref{eq:xi-1})
\begin{equation}
\delta P_{\rm down} = \mu_{\rm inner} \mu_{\rm outer} \sum_{\rm res} \sum_{lm} \Re \left[\alpha_{lm} \frac{\widetilde{Z}^{\rm down\ast}_{lm,\vec N_{\rm outer}}\widetilde{Z}^{\rm down}_{lm,\vec{N}_{\rm inner}}}{\omega^2}
e^{i(\vec N_{\rm outer}\cdot\vec\psi^{(0)}_{\rm outer} - \vec N_{\rm inner}\cdot\vec\psi^{(0)}_{\rm inner})}\right].
\end{equation}
This leads to the overall power absorption by the inner body:
\begin{equation}
P_{\rm abs}
= -\mu_{\rm inner}\mu_{\rm outer}
\sum_{\rm res} \sum_{lm} \Re \left[ \frac{
\widetilde{Z}^{\rm down\ast}_{lm,\vec N_{\rm outer}}
}{\omega^2}
\left(c_{13}^\ast
\widetilde{Z}^{\rm out}_{lm,\vec{N}_{\rm inner}} +
\alpha_{lm} \widetilde{Z}^{\rm down}_{lm,\vec{N}_{\rm inner}}\right)
e^{i\Theta}\right].
\label{eq: power_abs2}
\end{equation}
Thus the power absorbed can be calculated entirely from the radial components of $\psi_{4}$ with both contributions from the test particle's radiation and the perturbing body's radiation, without explicitly reconstructing the metric perturbations.

\subsection{Selection rules}
\label{ss:selection}

The selection rules for $Z^{\rm down,out}$ imply a set of selection rules for resonances. The azimuthal symmetry rule ($m=N_\phi$) implies that a resonance only contributes to Eq.~(\ref{eq: power_abs2}) if $m = N_{\phi,\rm inner} = N_{\phi,\rm outer}$. Furthermore, Eq.~(\ref{eq:equator}) implies that if we flip the signs of $m$, $\vec N_{\rm inner}$, and $\vec N_{\rm outer}$ (and hence $\omega$), and use that this flip implies $c_{13}\rightarrow c_{13}^\ast$ (Section 5.1 of Ref.~\cite{2011MNRAS.414.3212H}) and $\Theta\rightarrow-\Theta$, then each resonance term in the sum of Eq.~(\ref{eq: power_abs2}) is paired with a resonance of the opposite sign that differs by a complex conjugate and a factor of $(-1)^{l+N_{\theta,\rm inner}}(-1)^{l+N_{\theta,\rm outer}}$. The real part then cancels out unless $N_{\theta,\rm inner}$ and $N_{\theta,\rm outer}$ are both even or both odd. We thus have the resonance selection rules:
\begin{equation}
m = N_{\phi,\rm inner} = N_{\phi,\rm outer}
~~~{\rm and}~~~
N_{\theta,\rm inner}-N_{\theta,\rm outer} = {\rm even}~~~
({\rm selection~rules}).
\label{eq:selection}
\end{equation}
If the selection rules are obeyed, then the $m\rightarrow-m$, $\vec N_{\rm inner}\rightarrow-\vec N_{\rm inner}$, $\vec N_{\rm outer}\rightarrow-\vec N_{\rm outer}$ term in the sum is the complex conjugate, and so we may ignore the real part in Eq.~(\ref{eq: power_abs2}) without loss of generality.

If one set of $\vec N_{\rm inner},\vec N_{\rm outer}$ obeys the resonance condition and selection rules, then so will any integer multiple of these vectors. So we will choose the vector with smallest magnitude to be the ``fundamental'' resonance $\vec N_{\rm inner,F}, \vec N_{\rm outer,F},\omega_{\rm F}$ (subscript F), and then consider multiples of this as $s\vec N_{\rm inner,F}, s\vec N_{\rm outer,F}, s\omega_{\rm F}$, where $s = \pm 1,\pm 2, \pm 3 ...$ is a nonzero integer. Note that if $\vec N_{\rm inner,F}, \vec N_{\rm outer,F}$ satisfies the selection rules (Eq.~\ref{eq:selection}), then all of its multiples $s\vec N_{\rm inner,F}, s\vec N_{\rm outer,F}$ satisfy the selection rules as well.

\subsection{Putting it all together}

We can use Eq.~(\ref{eq: power_abs2}) for a single resonance to and the power to action derivative relation (Eq.~\ref{eq: P_abs_J_dot}) to give the expression for $\langle 
\dot{\tilde{J}}_{{\rm td},i} \rangle$:
\begin{equation}
\label{eq: J_dot_td_Pabs}
\langle 
\dot{\tilde{J}}_{{\rm td},i} \rangle_{\rm single\,resonance} = -\mu_{\rm outer} \sum_{l} \frac{N_{{\rm inner},i}\tilde{Z}^{\rm down\ast}_{\vec{N}_{\rm outer}, lm}}{\omega^3}\left(c^{*}_{13}\tilde{Z}^{\rm out}_{\vec{N}_{{\rm inner}}, lm} + \alpha_{ lm}\widetilde{Z}^{\rm down}_{\vec{N}_{\rm inner}, lm}\right) e^{i\Theta} .
\end{equation}
(There is no need for summation over $m$ since each resonance has a single value of $m$ according to the selection rules; and for resonances satisfying the selection rules, we may drop the real part.) If there is a fundamental resonance with mode vectors $\vec N_{\rm inner,F}$ and $\vec N_{\rm outer,F}$ then we have a summation:
\begin{equation}
\label{eq: J_dot_td_Pabs_s}
\langle 
\dot{\tilde{J}}_{{\rm td},i} \rangle_{{\rm multiples\,of\,F}} = -\mu_{\rm outer} \sum_{s=\pm1,\pm2,\pm3,...} \sum_{l} \frac{sN_{{\rm inner,F},i}\tilde{Z}^{\rm down\ast}_{s\vec{N}_{\rm outer,F}, lm}}{\omega^3}\left(c^{*}_{13}\tilde{Z}^{\rm out}_{s\vec{N}_{{\rm inner,F}}, lm} + \alpha_{ lm}\widetilde{Z}^{\rm down}_{s\vec{N}_{\rm inner,F}, lm}\right) e^{is\Theta_{\rm F}} ,
\end{equation}
where $m_{\rm F}$ is the azimuthal index of the fundamental resonance, the higher harmonics of the resonance have azimuthal index $m=sm_{\rm F}$ and resonant argument $\Theta=s\Theta_{\rm F}$, and the right-hand side is real because the $s$ term is the complex conjugate of the $-s$ term.

We can also rewrite Eq.~(\ref{eq: J_dot_td_Pabs_s}) in a similar form to Eq.~(\ref{eq: delta_J}) as
\begin{equation}
\tilde{G}_{\vec{N}_{\rm inner},\vec N_{\rm outer}, {\rm td},i}(\tilde{\vec{J}}_{\rm inner}, \tilde{\vec J}_{\rm outer})
= -\mu_{\rm outer} \sum_{l} \frac{N_{{\rm inner},i}\tilde{Z}^{\rm down\ast}_{\vec{N}_{\rm outer}, lm}}{\omega^3}\left(c^{*}_{13}\tilde{Z}^{\rm out}_{\vec{N}_{{\rm inner}}, lm} + \alpha_{ lm}\widetilde{Z}^{\rm down}_{\vec{N}_{\rm inner}, lm}\right).
\label{eq:G1}
\end{equation}
In the usual case where there is a single fundamental resonance, then using Eq.~(\ref{eq: delta_J_expand}) we may write the change in action as a sum over the overtone index $s$:
\begin{equation}
    \Delta \tilde{J}_{\rm inner,i} \approx
     \sum_{s=\pm 1,\pm2,\pm3,...}  \tilde{G}_{s \vec{N}_{\rm inner,F}, s\vec N_{\rm outer,F}, {\rm td},i}(\tilde{\vec{J}}_{\rm inner}, \tilde{\vec J}_{\rm outer}) e^{is\Theta_{\rm res,F}}
    e^{i\pi ({\rm sgn}\,s\Gamma)/4} \sqrt{\frac{2 \pi}{|s\Gamma|}}.
\label{eq:G2}
\end{equation}

Equations~(\ref{eq:G1}) and (\ref{eq:G2}) allow us to compute the change of the inner particle's action variable by specifying the resonance condition for the inner and outer orbits ($\vec N_{\rm inner,F}$ and $\vec N_{\rm outer,F}$), the angles at resonance crossing, and using an existing Teukolsky solver to determine the wave amplitudes $\tilde Z_{\vec N,lm}^{\rm out,down}$ and the scattering parameters $c_{13}$ and $\alpha_{lm}$. They constitute the main result of this paper.

\section{Examples}\label{sec: Testing_models}

Now that we have computed the change in the actions due to the external tidal field, Eq.~(\ref{eq: J_dot_td_Pabs_s}), we can compare our calculation to the results found in the literature, typically for the case of a static perturber. Our test cases will be: precession resonance for a Keplerian system and a generic tidal resonance for a static tidal field.  We will conclude this section with a comparative analysis for a tidal resonance crossing with a nearby perturber. We used the Teukolsky integrator developed in Ref.~\cite{2011MNRAS.414.3212H} to numerically compute the orbital frequencies $\Omega^i$ and GW amplitudes $\tilde Z^{\rm out,down}_{\vec N,lm}$, which was implemented in {\tt C}.

\subsection{Precession Resonance}
\label{ss:prec}

\begin{figure}
    \centering
    \includegraphics{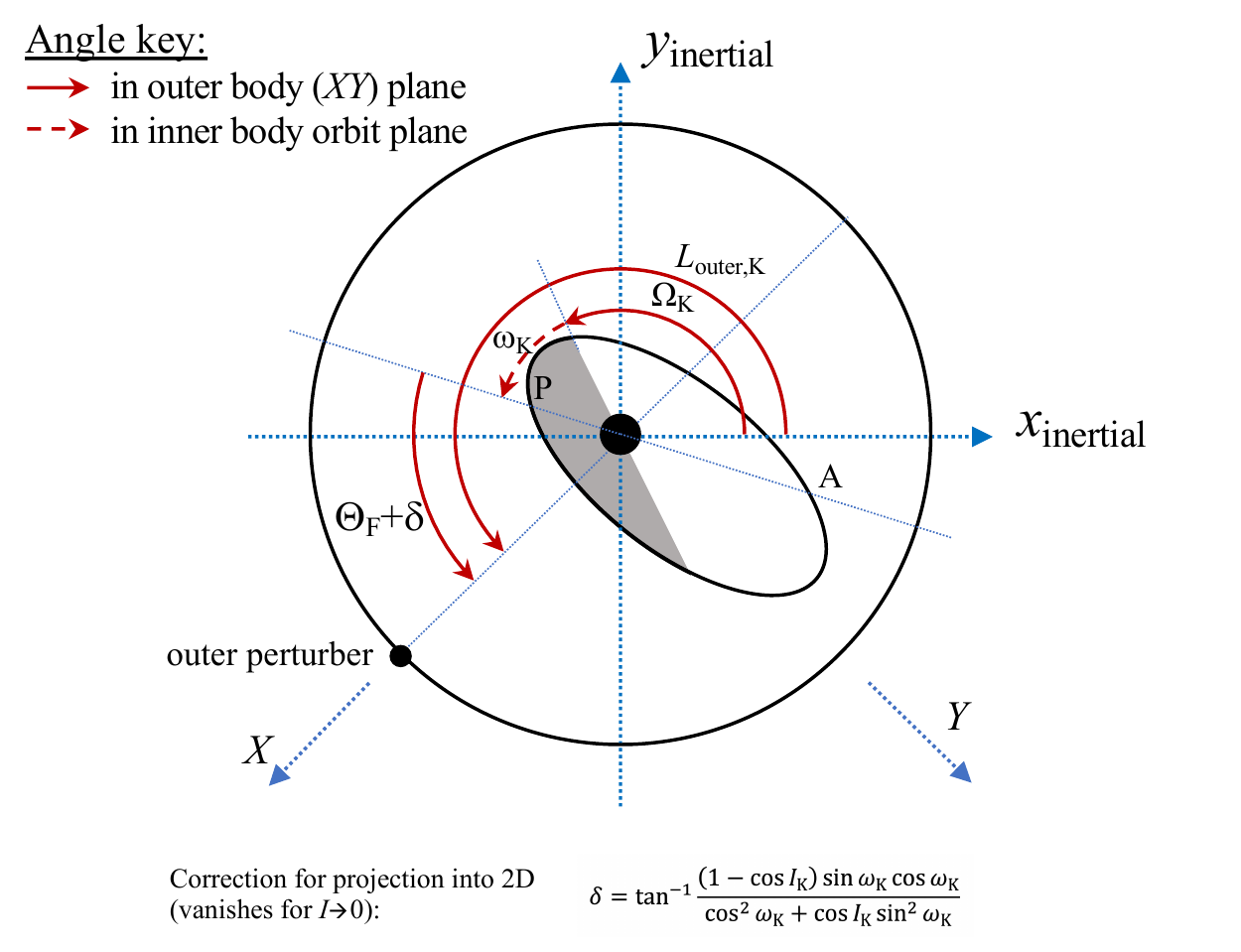}
    \caption{\label{fig:grprec}The setup for the precession resonance problem (Sec.~\ref{ss:prec}). The pericenter (P) and apocenter (A) of the inner body precess at approximately the same rate as the outer body orbits, so that the angle $\Theta_{\rm F}+\delta$ between the inner apsidal line (P--A) and the outer perturber is roughly constant. The shaded region shows the part of the inner body orbit that is ``above'' the plane of the page ($z_{\rm inertial}>0$). The angle $\delta$ is a small correction that oscillates around zero due to the argument of pericenter $\omega_{\rm K}$ not being in the same plane as the other angles. The figure is not to scale.}
\end{figure}

The first case we consider is the precession resonance described in Ref.~\citep{2022PhRvD.105b4017K}. This is the resonance between an inner orbit's apsidal precession rate with an outer orbit's orbital period. For this setup, the inner body is in a generic orbit, while the external perturbing body is on a circular equatorial orbit. The precession rate (rate of change of the pericenter around the black hole in the plane of the orbit) of the inner body is in resonance with the period of the outer body. Our main interest in this first case is to compare our resonant torque calculation to a case where Keplerian arguments are applicable. Specifically, will analyze an EMRI system where the outer body is near the Keplerian regime ($r_{outer} \sim 500M$) and the spacetime is that of a Schwarzschild black hole. We will compute the tidal torque on the inner body, $\langle \dot{\tilde{L}}_{\rm z} \rangle$, as a function of the resonant argument $\Theta_{\rm F}$, which is the mean angle between the apsidal line of the inner body and the direction to the outer perturber (see Fig.~\ref{fig:grprec}). We will compare this with the orbit averaged Keplerian torque which will be, at lowest order, quadrupolar since the torque is related to a second-order moment in position.

For the general precession resonance studied in Ref.~\cite{2022PhRvD.105b4017K}, the resonance condition is $q\dot{\omega}_K = pn_{\rm outer}$, where $\dot{\omega}_K$ is the rate of precession of the inner body's pericenter, $n_{\rm outer}$ is the mean motion of the perturbing body, and $p$ and $q$ are integers. In the quadrupole limit for a circular orbit, only the 2:2 resonance ($p=q=2$) survives; other resonances exist for an eccentric outer body. This resonant argument is $2\Theta_{\rm F}$, where
\begin{equation}
\Theta_{\rm F} = L_{K,\rm outer} - (\Omega_K+\omega_K)
\label{eq:theta-f-prec}
\end{equation}
is the fundamental resonance argument expressed in terms of Keplerian orbital elements.

\subsubsection{Keplerian calculation}

We can compare our computed torque with the Keplerian expression, which should be valid in the non-relativistic limit. (See Ref.~\cite{1999ssd..book.....M} for an extensive discussion of these calculations for Keplerian orbits.) We will furthermore make the quadrupole order approximation (the leading order term in $r_{\rm inner}/r_{\rm outer}$). We build a capital Roman letter ($XYZ$) Cartesian coordinate system placing the $Z$-axis along the angular momentum vector of the outer body, and the $X$-axis in the direction from the black hole to the outer body. (This frame will rotate at an angular velocity $\Omega^\phi_{\rm outer}$.)
The quadrupole tidal potential is given by $\Phi_{\rm tidal}=(\mu_{\rm outer}/2r_{\rm outer}^3)(Y^2+Z^2-2X^2)$.
In Newtonian gravity, the torque on the inner particle per unit mass due a tidal field averaged over an orbit of the inner particle, $\dot{\tilde{L}}_{\rm z}$, is given by
\begin{align}
\label{eq: kepler_torque}
\langle \dot{\tilde{L}}_{\rm z} \rangle 
= \left\langle -X\frac{\partial\Phi_{\rm tidal}}{\partial Y} + Y\frac{\partial\Phi_{\rm tidal}}{\partial X} \right\rangle
= -\frac{3\mu_{\rm outer}}{r_{\rm outer}^3}\langle XY \rangle, ~~~ \langle ... \rangle \equiv \int^{2\pi}_{0}\frac{dM_K}{2 \pi} ...,
\end{align}
where the $\langle ... \rangle$ is an average over the mean anomaly, $M_K$, of the inner orbit \citep{1999ssd..book.....M}. To simplify this computation, we compute this second moment in position in the rotated frame of the inner orbit (lower-case Roman letters: $xyz$, with $z$ along the inner orbit angular momentum vector and $x$ pointing from the black hole to the perigee of the inner orbit), and then rotate back to the original frame using the general Euler rotations:
\begin{align}
\label{eq: moment_rotate_unrotate}
\langle r^I r^J\rangle = R^I{_i}R^J{_j}\langle r^i r^j \rangle,
\end{align}
where the $R^I{_i}$ is the matrix that rotates from the $xyz$ frame to the $XYZ$ frame (Euler rotations) defined by the Keplerian orbital elements $\omega_K$, $I_K$, and $\Omega_K$ \citep{1999ssd..book.....M}.
We can expand our expression for the torque to yield
\begin{eqnarray}
\label{eq: moment_unrotate}
\langle \dot{\tilde{L}}_{\rm z} \rangle
&=& -\frac{3\mu_{\rm outer}}{r_{\rm outer}^3}((\cos{\Omega_K}\cos{\omega_K} - \sin{\omega_K}\sin{\Omega_K}\cos{I_K})(\cos{\omega_K}\sin{\Omega_K} + \sin{\omega_K}\cos{\Omega_K}\cos{I_K})\langle x^2 \rangle \nonumber
\\
&&
+ (\sin{\omega_K}\cos{\Omega_K} + \cos{\omega_K}\sin{\Omega_K}\cos{I_K})(\sin{\omega_K}\sin{\Omega_K} - \cos{\omega_K}\cos{\Omega_K}\cos{I_K})\langle y^2 \rangle),
\end{eqnarray}
where $a$ and $e$ are the semi-major axis and the eccentricity, respectively, and $\langle x^2\rangle$ and $\langle y^2 \rangle$ are the average of the inner body's quadrupole moments in its orbit plane, given by
\begin{equation}
\langle x^2 \rangle = a^2\frac{1+4e^2}{2} ~~{\rm and}~~ \langle y^2 \rangle = a^2\frac{1-e^2}{2}.
\end{equation}
(Note that other moments such as $\langle xy\rangle$ are zero.)

The fundamental precession resonant argument is $\Theta_{\rm F} = \psi^\phi_{\rm outer} + \psi^r_{\rm inner} - \psi^\phi_{\rm inner}$. Expressed in the usual Keplerian orbital elements, we have $\psi^\phi_{\rm outer} = L_{K, \rm outer}$ (the mean longitude), $\psi^r_{\rm inner} = M_K$ (the mean anomaly), and $\psi^\phi_{\rm inner} = \Omega_K + \omega_K + M_K$. Since we referenced the Keplerian orbital elements of the inner body relative to the $XYZ$ frame, which places the $X$-axis pointing to the outer body, the $\Omega_K - L_{K,\rm outer}$ for a general $X$-axis in the outer body's orbital plane becomes $\Omega_K$ in our calculation, and so the fundamental argument (Eq.~\ref{eq:theta-f-prec}) becomes
$\Theta_{\rm F} = - (\Omega_K + \omega_K)$.
Since we only consider terms that satisfy our precession resonance condition, as all other angles will get time-averaged out, we take the Fourier expansion of Eq.~(\ref{eq: moment_unrotate}) in $\Omega_K$ and $\omega_K$, and keep only terms whose arguments are multiples of $\Omega_K+\omega_K$. We thus arrive at the final expression for $\langle \dot{\tilde{L}}_{\rm z} \rangle$:
\begin{eqnarray}
    \langle \dot{\tilde{L}}_{\rm z} \rangle_{\rm resonant} &=& -\frac{3 \mu_{outer}}{8r_{outer}^3}(\langle x^2 \rangle - \langle y^2 \rangle)(1 + \cos{I_K})^2 \sin 2(\Omega_K+\omega_K)
\nonumber \\
&\rightarrow& 
\frac{3 \mu_{outer}}{8r_{outer}^3}(\langle x^2 \rangle - \langle y^2 \rangle)(1 + \cos{I_K})^2 \sin 2\Theta_{\rm F}
\nonumber \\
&=& 
\frac{15 \mu_{outer}}{16r_{outer}^3}a^2e^2(1 + \cos{I_K})^2 \sin 2\Theta_{\rm F}.
\label{eq: final_keplerian_torque}
\end{eqnarray}
We can see that Eq.~(\ref{eq: final_keplerian_torque}) depends on twice our fundamental resonant angle. This tells us that when  comparing our torque with the Keplerian case, we need to start at the $s=\pm 2$ overtone in order to see this quadrupolar effect. 

\begin{figure}[ht]
    \centering
    \includegraphics[scale=0.9]{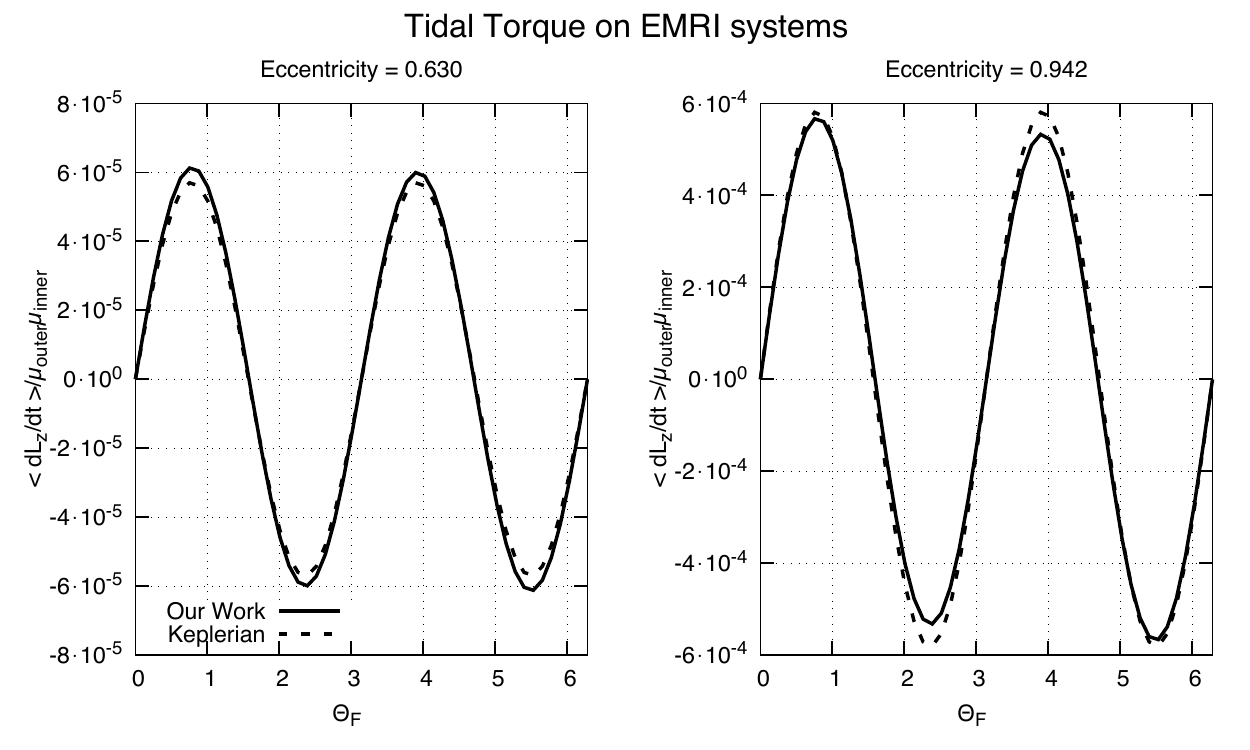}
    \caption{The plots shows the tidal torque ($ \langle \dot{\tilde{L}}_{\rm z} \rangle$) on an EMRI due to an effectively static perturber on a circular orbit of radius $r_{\rm outer} = 500M$. The left (right) plot is for an EMRI orbit with eccentricity $e = 0.63$ (0.942), pericenter $r_p = 30M$ ($10M$) and inclination $\theta_{\rm inc} = \pi/4$. We compare the Keplerian (dashed) and our torques and find about an $8\%$ difference in amplitude.} 
    \label{fig:multiplot_keplerian}
\end{figure}

The results for two EMRI cases with different inner body eccentricities are shown in Fig.~\ref{fig:multiplot_keplerian}. The numerical parameters used are the range of overtone index $|s|_{\rm max} = 2$, number of $\ell$-modes $n_{\ell} = 4$, outer radius $r_{\rm outer} = 500M$, pericenter $r_{\rm peri} = (10, 30)$, apocenter $r_{\rm apo} = (335,132)$, mass of the cBH $M = 1$, and spin parameter $a_{*} = 0$.

\subsubsection{Comparison with our fully relativistic calculation}

We can see from Fig.~\ref{fig:multiplot_keplerian}, that for smaller eccentricities, the Keplerian expression for the torque is in stronger agreement with our work. The largest difference in maxima of $|\dot{\tilde L}_z|$ between our work and the Keplarian formulation are 7.5\% and 8.3\% for the $e_K=0.630$ and $e_K=0.942$ caess, respectively. This makes sense since the Keplerian expression is a quadrupolar approximation and would require higher multipole moments for orbits with greater eccentricities. There is also the subtle behaviour that the peaks of our results differ slightly from a perfect sinusoid. This is because the sum performed resonance overtones, $s$, starts at $-2$ then proceeds to $+2$ in integer steps (skipping $s=0$ since this contributes no power) and it is the $s=\pm 1$ contribution that causes a slight increase/decrease in the amplitude of our work as we sweep across $\Theta_{\rm F}$ since those terms are proportional to $\sin\Theta_{\rm F}$. This subtle effect is not captured the by quadrupole moment expression alone and would require the next order corrections -- the octupole ($\ell=3$, $|m|=1$) and the gravitomagnetic quadrupole ($\ell=2$, $|m|=1$) perturbations. These are all consistently captured, with the fully relativistic effects, by our analysis using the Teukolsky equation.

\subsection{Tidal resonance from distant perturber}

We can also test our code by comparison to the cases in Refs.~\citep{2019PhRvL.123j1103B, 2021PhRvD.104d4056G,  2022arXiv220504808G}, where we will look for the changes in the action variables of an inner body due to the tidal influence of a distant outer body. These analyses have been done in the limit of the outer body being both distant and static, but with a rotating black hole and an inner body at general $r/M$. They therefore represent a good comparison case for our work. Tidal resonance occurs when some integer linear combination of the fundamental frequencies of the inner body is close to zero and thus can be resonant with the much slower motion of the outer body (see Eq.~\ref{eq:resonance_condition}). Since Refs.~\citep{2019PhRvL.123j1103B, 2021PhRvD.104d4056G, 2022arXiv220504808G} treated a zero-frequency tidal field, we will approximate their setup by putting the outer body very far away, e.g., $r_{\rm outer}=500M$.
We will compare our torque for the same resonance cases analyzed in Ref.~\cite{2019PhRvL.123j1103B, 2021PhRvD.104d4056G, 2022arXiv220504808G}, which correspond to the case of Eq. (\ref{eq:resonance_condition}) with $(n,k,m) = (-3,0,2)$ and $(3,0,-2)$ with different initial orbit conditions for each case.

\begin{figure}[ht]
    \centering
    \includegraphics[scale=0.9]{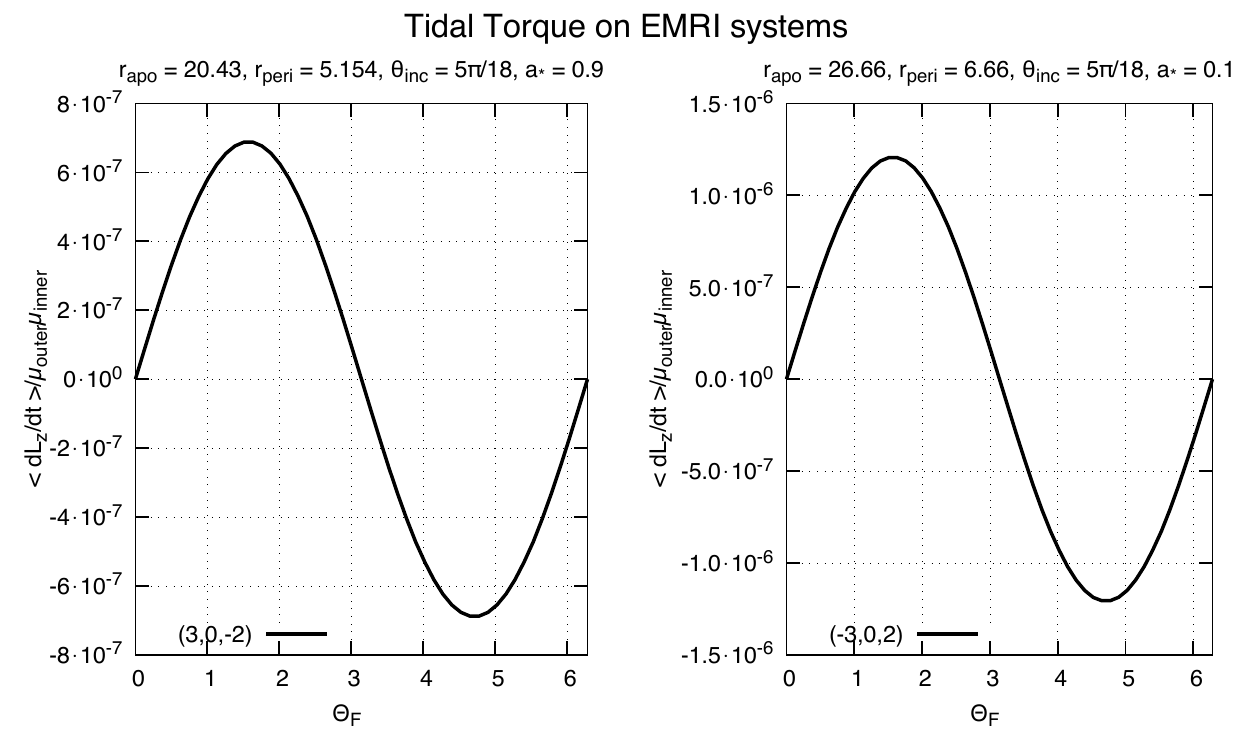}
    \caption{This shows the tidal torque on the inner body at different tidal resonance conditions, $\vec{N}_{\rm inner} = (3,0,-2)$ and $(-3,0,2)$, and orbit parameters (shown in each panel). The radius of the outer perturber is $r_{\rm outer} = 500M$, effectively making the perturber ``static.''} 
    \label{fig:multiplot_tidalres}
\end{figure}

In our analysis, we computed $\langle \dot{\tilde{J}}_{\rm td,\phi} \rangle$, where our overdot is a derivative with respect to the coordinate time, $t$, from Eq.~(\ref{eq: J_dot_td_Pabs}) with the index set to the $\phi$-coordinate. This differs from our compared Ref.~\cite{2019PhRvL.123j1103B}, which computed the evolution with respect to inner body proper time, though we compute the relative difference between either evolution parameter using the redshift invariant in Appendix~\ref{app:time}. This allows us to compare the ``static'' tidal field limit in Ref.~\citep{2019PhRvL.123j1103B} with our general calculation. Figure~\ref{fig:multiplot_tidalres} shows the tidal torque due to a perturber in the static limit of an EMRI system. The numerical parameters for these plots are: the number of $\ell$-modes $n_{\ell} = 4$; maximum overtone index $|s|_{\rm max} = 2$; mass of the black hole $M = 1$ (which is merely a choice of units); and the apocenter, pericenter, inclination, and spin parameter ($a_{*}$) for each case of $\vec{N}_{\rm inner}$ are given on top of the plots.

In Fig.~\ref{fig:multiplot_tidalres}, the amplitude of both plots are, in order of left to right, are about $6.9 \times 10^{-7}$ and $1.20 \times 10^{-6}$. The case when $(n,k,m) = (3,0,-2)$, Ref.~\citep{2021PhRvD.104d4056G} quote a value of the tidal torque of about $40e^2/(e-1)^2/r^3_{\rm outer}$ from their Fig.~4 (with the same orbit and BH parameters), which for an eccentricity of $e=0.62$ and $r_{\rm outer} = 500M$, corresponds to $8.5\times 10^{-7}$, about $19\%$ difference from our amplitude. If we increase $r_{\rm outer}$ to $1000M$, our computed amplitude becomes $9.24\times 10^{-8}$, corresponding to a $13\%$ difference from the amplitude quoted in Ref.~\citep{2021PhRvD.104d4056G} at the same $r_{\rm outer}$; thus the convergence to the static case for increasing $r_{\rm outer}$ appears to be slower for this set of resonance and physical parameters. Slower convergence of resonant strengths to the large-$r_{\rm outer}$ limit has also been seen for other resonance problems \cite{2011MNRAS.414.3212H}. Similarly for the case of $\vec{N}_{\rm inner} = (-3,0,2)$, we can compare this amplitude to Fig.~4 of Ref.~\citep{2022arXiv220504808G} at the zero inclination case for the perturber. They quote a value for the tidal torque of $150/r^3_{\rm outer}$, which for our case where $r_{\rm outer} = 500M$, corresponds to $\dot L_z /\mu_{\rm inner}\mu_{\rm outer} = 1.2 \times 10^{-6}$. 

\begin{figure}[ht]
    \centering
    \includegraphics[scale=0.9]{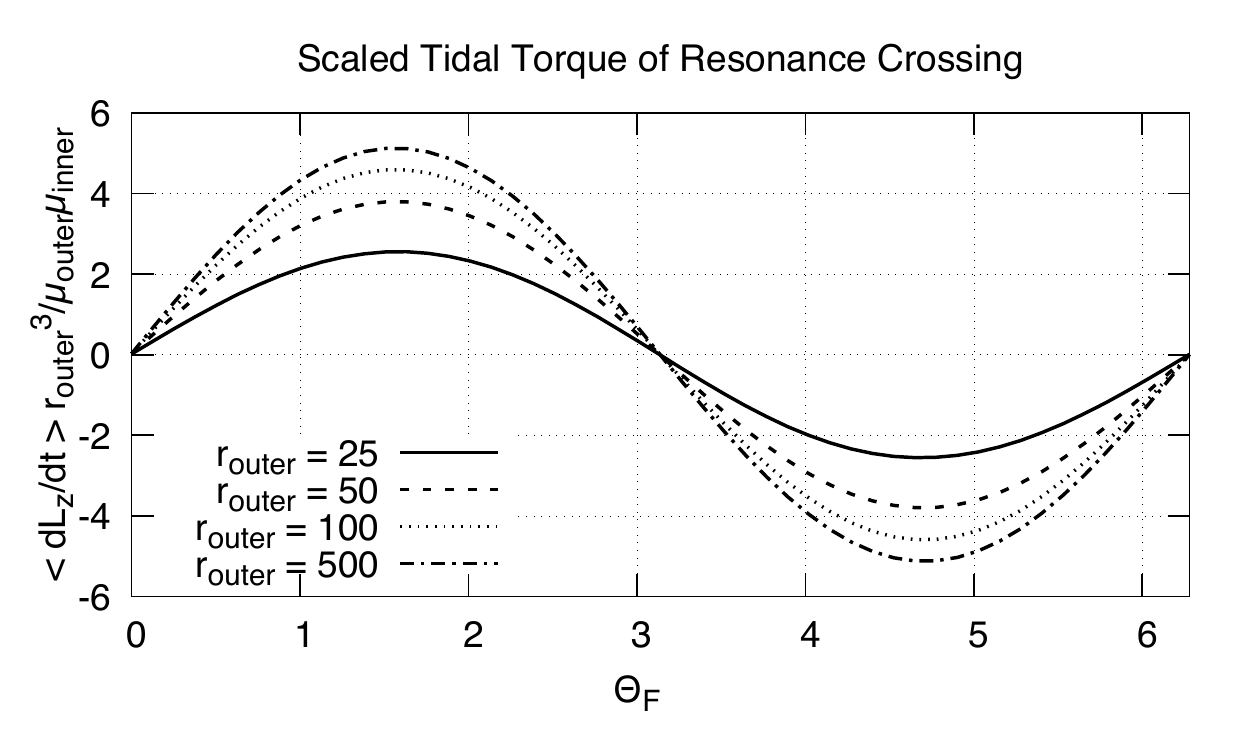}
    \caption{The tidal torques per unit mass of the inner and outer bodies with the respective $1/r^3_{\rm outer}$ scaled out for the mode vector $\vec{N}_{\rm inner} = (1,2,-2)$ and for a pericenter, inclination, and black hole mass and spin of $3M$, $\pi/4$, $M=1$, and $a_{*} = 0.9$, respectively The apocenter changes according to the outer orbit radius, decreasing as the outer radius decreases. Each curve represents the scaled torques on the EMRI system due to a perturbing orbit at a specific radius. Regardless of the outer orbit radius, we can see a similar behavior with the tidal torques on the EMRI.} 
    \label{fig:tidal_res_scaled}
\end{figure}

Finally, we have explored the dependence on the radius of the outer body orbit. We also plotted the tidal torques with the outer radius scaled out in Fig.~\ref{fig:tidal_res_scaled} for the $\vec N_{\rm inner}=(1,2,-2)$ resonance \cite{2019PhRvL.123j1103B} with fixed pericenter $r_{\rm peri}=3M$, inclination $I=\pi/4$, black hole mass $M=1$ and spin $a_\star = 0.9$, but for different radii of the perturbing body. Table~\ref{table: tidal_scaled_data} displays the values of the torques in Fig.~\ref{fig:tidal_res_scaled} at $\Theta_{\rm F}=0$ and $\Theta_{\rm F} = 1.634$, which is close to the first maxima. As the outer body moves closer in, the strength of the torque increases, which we expect since resonant interactions are expected to increase for closer perturbers, but there are deviations from the power law of $\propto r^{-3}_{\rm outer}$ that we expect for a distant perturber in the Keplerian regime, as evident in the data displayed in Table~\ref{table: tidal_scaled_data} at different values of $r_{\rm outer}$.

\begin{table}[ht]
\begin{tabular}{|c|c|l|c|c|}
\hline\hline
apocenter & $r_{\rm outer}$ & ~~$\Theta_{\rm F}$ & $\langle \dot{{L}}_{\rm z} \rangle /\mu_{\rm inner} \mu_{\rm outer}$ & $\langle \dot{{L}}_{\rm z} \rangle r^3_{\rm outer} /\mu_{\rm inner} \mu_{\rm outer}$ \\ \hline
\multirow{2}{*}{4.75$M$} & \multirow{2}{*}{25$M$} & 0 & $1.072 \times 10^{-6}$ & $1.675 \times 10^{-2}$ \\
& & 1.634 & $1.633 \times 10^{-4}$ & 2.552 \\ \hline
\multirow{2}{*}{5.67$M$} & \multirow{2}{*}{50$M$} & 0 & $1.666 \times 10^{-7}$ & $2.082 \times 10^{-2}$ \\
& & 1.634 & $3.034 \times 10^{-5}$ & 3.792 \\ \hline
\multirow{2}{*}{6.19$M$} & \multirow{2}{*}{100$M$} & 0 & $2.280 \times 10^{-8}$ & $2.280 \times 10^{-2}$ \\
& & 1.634 & $4.583 \times 10^{-6}$ & 4.583 \\ \hline
\multirow{2}{*}{6.52$M$} & \multirow{2}{*}{500$M$} & 0 & $1.900 \times 10^{-10}$ & $2.375 \times 10^{-2}$ \\
& & 1.634 & $4.091 \times 10^{-8}$ & 5.113 \\ \hline\hline
\end{tabular}
\caption{Table of values in Fig.~\ref{fig:tidal_res_scaled} at $\Theta_{\rm F} = 0$ and 1.634, where $M = 1$. We show the torques both with and without extracting the usual $\propto 1/r_{\rm outer}^3$ scaling for a tidal field.}
\label{table: tidal_scaled_data}
\end{table}

In principle, the resonant torque as a function of $\Theta_{\rm F}$ could be maximum at any phase, and one must study the system in detail to know whether to expect a cosinelike dependence in Fig.~\ref{fig:multiplot_tidalres}, a sinelike dependence, or a general linear combination. In this case, we expect to see an approximate $\sin\Theta_{\rm F}$ dependence in the limit of a static tidal field of the perturber on the Kerr spacetime. The physical reason behind this lies in the discrete symmetry of the spinning black hole and the perturbation, that is the combination of longitude reversal $\phi \rightarrow -\phi$ and time reversal $t \rightarrow -t$. This symmetry is actually broken for the tidally perturbed Kerr spacetime because the absorbing boundary condition at the horizon does not respect the symmetry. (Recall that a tidal field that is static in the inertial frame has a frequency $-m\Omega_{\rm H}$ in the frame rotating with the horizon at angular velocity $\Omega_{\rm H}$.) This manifests itself as an imaginary (dissipative) part to the Love numbers $k_{2m}$ \citep{2021PhRvL.126m1102L, 2021PhRvD.103h4021L, 2021JHEP...05..038C} for $m\neq 0$. However, the imaginary part of the Love number is numerically very small, $\Im k_{22} = -\frac1{10}a_\star(a_\star^2+\frac13)$ when referenced to a radius of $2M$ (see Eq.~5.15 of Ref.~\citep{2021JHEP...05..038C}).\footnote{The Love number is dimensionless and is always referenced to some radius $r_s$; if another radius is chosen, it scales as $\propto r_s^{-5}$.}
We thus expect an {\em approximate} $t\rightarrow -t$, $\phi\rightarrow -\phi$ symmetry, particularly for orbits that stay well outside of $r = 2M$.
We can see from the expressions for the torque in Ref.~\cite{2019PhRvL.123j1103B} that under these transformations, $\dot{\tilde{L}}_{\rm z} \rightarrow - \dot{\tilde{L}}_{\rm z}$ and $\Theta_{\rm F} \rightarrow -\Theta_{\rm F}$. For an orbit at $\Theta_{\rm F} = 0$, the time and azimuthal transformations give $\langle  \dot{\tilde{L}}_{\rm z} \rangle = - \langle  \dot{\tilde{L}}_{\rm z} \rangle $. This implies that, in the static tidal field limit, the $\Theta_{\rm F}$ dependence would be dominated by the $\sin\Theta_{\rm F}$ dependence, though we also see from Fig.~\ref{fig:tidal_res_scaled} this strong sine dependence for perturber orbits as close as $25M$ from the black hole.

\section{Discussion}
\label{section: discussion}

Our completely relativistic approach to computing the change in action variables due to tidal fields, $\langle \dot{\tilde{J}}_{\rm td, i} \rangle$, provides a new and efficient method for computing resonant interactions in three body EMRI problems and is in agreement with the previously studied static and Keplerian limiting cases \cite{2019PhRvL.123j1103B, 2022PhRvD.105b4017K, 2022arXiv220504808G}. We can see from Figs.~\ref{fig:multiplot_keplerian} and \ref{fig:multiplot_tidalres}, that our calculation of $\langle \dot{\tilde{J}}_{\rm td,i} \rangle$ converges well to the limiting cases for static perturbations under various resonance conditions as we increase the number of modes, $n_{\ell}$, and the overtone parameter, $s$. 

This technique has the advantage tracking the dynamics of an inspiral in the strong gravity regime while not having to explicitly compute metric perturbations in order to analytically express the evolution of the action variables and, therefore, an efficient approach to calculating waveforms that result from gravitational perturbations on Kerr spacetimes. We achieve this by using the invariant Weyl scalar, $\psi_4$, which encodes all the gravitational wave data going down and away from the black hole. This invariant scalar (derived as a solution to the Teukolsky equations for gravitational perturbations) \citep{1973ApJ...185..635T} allows us to relate the power exchanged during a resonance crossing to the evolution of the action variables.

By using the invariant Weyl scalar, $\psi_4$, to compute the transfer of energy during a resonance crossing, we can relate this energy transfer between the two orbiting bodies to the change in the inner body's action variables induced by the gravitational perturbations of the external body. Since a significant change of the action variables occur only near a resonance crossing (since that is a stationary point in the orbit average), we only need to compute the $\psi_{4}$ of both bodies near that resonance crossing. We can then replace the perturbing body with the equivalent $\psi_{4}$ that satisfies the same boundary conditions of the physical body. This allows us to convert and solve a physical three body problem into an effective two body + incident gravitational wave problem \citep{2011MNRAS.414.3212H}.

Future work for this analysis could extend the perturbing body's orbit to be generic, like the EMRI. This would require the locations of the multiple resonance crossings and calculating the strength of each crossing. Another extension could be taking into account the intrinsic spin of the EMRI and perturber, as there could be resonance effects that couple the spin during a resonance crossing \cite{2022PhRvD.105l4040D, 2022PhRvD.105l4041D}. This formalism could also be used to analyze the case of perturbations from unbounded trajectories near the central black hole (parabolic encounters) by allowing the external $\psi_4$ to be a continuous function of frequency rather than discrete modes \citep{1977ApJ...213..183P,2010ApJ...725..353D}. Similarly to our analysis for computing $\Delta \tilde{J}_{\rm i}$, we could investigate computing the changes in the angle variables, $\Delta \psi^{\rm i}$, as this would allow for a complete Hamiltonian description of a relativistic three body resonance. Our technique could also be applied to the effect of a resonance crossing on a gravitational waveform, similar to what was done in Ref.~\citep{2019PhRvL.123j1103B} in their Eq.~(14) but using our expression Eq.~(\ref{eq:G2}) when integrating over the change in the phase. 

As discussed in prior works Refs.~\citep{2019PhRvL.123j1103B, 2021PhRvD.104d4056G, 2022arXiv220504808G}, resonances from external perturbers may play an important role in the gravitational waveforms from EMRI systems. The technique presented allows us to compute the effects on the emitted GWs from an EMRI during a resonance crossing from both a static and dynamic perturbation, i.e., the perturber can either be near or far from the EMRI; thus creating a more realistic analysis of galactic center environments and developing our understanding of gravitational perturbations on Kerr spacetimes. With the next generation of space-based interferometers coming online in the coming decades, such as LISA, we will be able to empirically probe these relativistic multi-body systems and will need a strong theoretical understanding of the possible dynamics within these strong gravity environments.

\begin{acknowledgments}

Mahalo nui loa to Kaimi Kahihikolo and Bryan Yamashiro for extensive debugging assistance and figure editing. And mahalo nui to Priti Gupta and B\'eatrice Bonga for insightful discussions and feedback on our analytic comparisons; and Matthew Digman and Xiao Fang for comments on the draft.

During the preparation of this work, M.S. and C.M.H. was supported by NASA award 15-WFIRST15-0008, Simons Foundation award 60052667, and the David \& Lucile Packard Foundation. 

Several figures in this paper were made with \textit{Gnuplot} \citep{gnuplot}.

\end{acknowledgments}

\appendix

\section{Proper time versus coordinate time}
\label{app:time}

In our analysis, we wrote our Hamiltonian with respect to the Boyer-Lindquist coordinate time, $t$, which differs from most other approaches to BHPT (e.g., \cite{2008PhRvD..78f4028H, 2013arXiv1305.5720E}) where proper time $\tau$ is used as the independent variable. This will cause our orbit-averaged time derivatives to differ from these other approaches by a factor of $\langle {d\tau}/{dt} \rangle_{T^3} = \langle 1/u^t \rangle_{T^3}$, where the bracket indicates averaging over the 3-torus of angles: $\langle \rangle_{T^3} = \int_{T^3} d^3{\vec\psi}/(2\pi)^3$.\footnote{This quantity is the natural generalization to a generic orbit of the redshift invariant, $1/\langle U\rangle_\tau$ \cite{2008PhRvD..77l4026D, 2011PhRvD..83h4023B}.}
We calculate this quantity in terms of action variables by working with a simple invariant of GR, the square of the inner body's 4-velocity: $u^{\alpha}u_{\alpha} = -1$. We can expand this out as:
\begin{equation}
u^t u_t + u^i u_i = u^\alpha u_\alpha = -1 ~~~ \rightarrow ~~~
\tilde{\mathcal{E}} - \frac{u^i}{u^t} u_i = \frac{1}{u^t},
\end{equation}
where we have multiplied by $-1/u^t$ and used the fact that $\tilde{\mathcal{E}} = -u_t$. Then we may take the 3-torus average:
\begin{equation}
\label{eq: dtau_dt}
    \tilde{\mathcal{E}} - \int_{T^3}\frac{d^3\psi}{(2\pi)^3}\frac{d \psi^j}{dt} \frac{\partial x^i}{\partial \psi^j} u_i = \int_{T^3}\frac{d^3\psi}{(2\pi)^3}\frac{1}{u^t} = \left \langle \frac{d\tau}{dt} \right \rangle,
\end{equation}
where we used the fact that $\tilde{\mathcal E}$ is constant over the 3-torus since it depends only on the actions;
we used the chain rule to rewrite the 3-velocity $u^i/u^t=dx^i/dt$ in terms of the angular frequencies (${d\psi^j}/{dt} = \Omega^j$); and at the end we used the definition of the 3-torus average.

Now recall the definition of the action coordinate:
\begin{align}
\label{eq: action_def}
    \tilde{J_{j}} = \frac{1}{2\pi}\oint_{C}u_{i}\, dx^{i},
\end{align}
where $C$ is some closed path around the invariant 3-torus where $\psi^j$ is incremented by $2\pi$ as one goes around the loop.
For any stationary Hamiltonian, the action coordinates are independent of the angle variables. Thus, we can average Eq.~(\ref{eq: action_def}) over the angle variables of the torus and its value will not change:
\begin{equation}
\label{eq: action_avg}
    \tilde{J_{j}} = \frac1{2\pi} \int_0^{2\pi} \frac{ d \psi^{k}}{2\pi} \int_0^{2\pi}  \frac{ d \psi^{l}}{2\pi} \int_0^{2\pi} u_{i} \frac{\partial x^i}{\partial \psi^j} d \psi^{j}
     = \int_{T^3} u_{i} \frac{\partial x^i}{\partial \psi^j} \frac{d^3 \psi}{(2\pi)^3},
\end{equation}
where in the first step we averaged over $\psi^k$ and $\psi^l$, and in the second step we specialized to the case where $(j,k,l)$ is a permutation of $(r,\theta,\phi)$; 
this allows us to cover the entire torus. 
Now we can simplify Eq. (\ref{eq: dtau_dt}) to give:
\begin{align}
\label{eq: fina_dtau_dt}
    \left \langle \frac{d\tau}{dt} \right \rangle = \tilde{\mathcal{E}} - \Omega^j \tilde{J}_j.
\end{align}

\nocite{*}

\bibliography{main}

\end{document}